\documentclass{article}
\usepackage{spconf,amsmath,graphicx}
\usepackage[paper=letterpaper,margin=1in]{geometry}
\newgeometry{top=1in,bottom=0.75in,right=0.75in,left=0.75in}

\usepackage{color}
\usepackage{amsthm}
\usepackage{mathtools}

\usepackage{bbm}
\usepackage{amssymb}
\usepackage{tikz}
\usepackage{enumitem}
\usepackage[algo2e,ruled,vlined,linesnumbered]{algorithm2e}
\SetInd{0em}{0.5em}
\SetKwComment{Comment}{/\!\!/}{}
\usepackage{booktabs}
\usepackage{cite}

\newcommand\blfootnote[1]{%
  \begingroup
  \renewcommand\thefootnote{}\footnote{#1}%
  \addtocounter{footnote}{-1}%
  \endgroup
}

\theoremstyle{definition}
\newtheorem{Example}{Example}

\usepackage{caption}
\begin{document}
\ninept
\title{On Throughput of Millimeter Wave MIMO Systems with Low Resolution ADCs}

\name{Abbas Khalili$^\dagger$, Shahram Shahsavari$^\ddagger$,  Farhad Shirani$^\dagger$, Elza Erkip$^\dagger$, Yonina C. Eldar$^\diamond$}
 \address{ $^\dagger$NYU Tandon School of Engineering, $^\ddagger$University of Waterloo, $^\diamond$ Weizmann Institute of Science}

\maketitle
\begin{abstract}
Use of low resolution analog to digital converters (ADCs) is an effective way to reduce the high power consumption of millimeter wave (mmWave) receivers. In this paper, a receiver with low resolution ADCs based on adaptive thresholds is considered in downlink mmWave communications in which the channel state information is not known a-priori and acquired through channel estimation.
A performance comparison of low-complexity algorithms for power and ADC allocation among transmit and receive terminals, respectively, is provided. Through simulation of practical mmWave cellular networks, it is shown that the use of low resolution ADCs does not significantly degrade the system throughput (as compared to a conventional fully digital high resolution receiver) when using the adaptive threshold receiver in conjunction with simple power and ADC allocation strategies.

 \blfootnote{This work is supported by National Science Foundation grant SpecEES-1824434 and NYU WIRELESS Industrial Affiliates. This work was done when S. Shahsavari was at NYU.}
\end{abstract}

\vspace*{-0.65cm}
\section{Introduction}
\vspace*{-0.15cm}
Millimeter wave (mmWave) systems have emerged as a promising candidate for high data rate communication in 5G wireless networks. One of the major obstacles in the implementation of mmWave systems is the high energy consumption \cite{rangan2014millimeter,walden1999analog,Murmann2015}. One way to reduce power consumption in mmWave systems is to use low resolution analog to digital converters (ADCs) (e.g. one-bit threshold ADCs) at the receiver \cite{MIMO1,mo2015capacity,alkhateeb2014mimo,abbasISIT2018,rini2017generalITW,mo2016ADC,dutta2020capacity,mezghani2012capacity,Dutta2019}. However, this inflicts a rate-loss due to the large quantization noise caused by coarse quantization.

There has been a large body of work dedicated to characterizing the capacity of point-to-point (PtP) MIMO systems in the presence of low resolution ADCs at the receiver
\cite{abbasISIT2018,rini2017generalITW,mo2016ADC,dutta2020capacity}. These works consider \textit{`analog-one-shot'} receivers, where at each channel-use the received signal goes through analog processing prior to being fed to the one-bit ADCs. The receiver then performs blockwise signal processing on the stored digital signal to decode the message. In contrast, \cite{abbasPtPISIT2019} and \cite{abbasMtISIT2019} propose two new classes of receivers with low resolution ADCs, called \textit{analog-blockwise} and \textit{adaptive threshold} receiver, respectively, which generalize analog-one-shot receivers and achieve higher performance in terms of communication rates for a given set of one-bit ADCs. These receivers incorporate delay elements to perform analog blockwise processing which is not possible with analog-one-shot receivers. More specifically, the adaptive threshold receiver changes the threshold of the ADCs adaptively based on their outputs in previous channel uses. Note that receivers with successive approximate register (SAR) ADCs, used for low power consumption applications \cite{5746277,5711005,5433830,6043594}, also belong to the family of adaptive threshold receivers.

A fundamental question which arises in the context of low resolution receivers is the best way to allocate a total of $m$ bits among the receiver antennas in order to maximize the achievable rate. Unlike analog-one-shot receivers which require pre-set ADCs of different resolutions for bit allocation among the antennas, the adaptive threshold receiver can form $m$-bit quantization using $m$ one-bit ADCs and allocate the bits to the antennas in any desired fashion.
This flexibility allows the receiver to switch between tasks which require different bit allocations among antennas such as channel estimation and data communication, where the latter could depend on the estimated channel. Another advantage of the adaptive threshold receiver is its  optimality (in terms of achievable rates) in the high SNR regime for the single and multi-user uplink (UL) and downlink (DL) communication scenarios \cite{abbasMtISIT2019}. The proposed transmission schemes in \cite{abbasMtISIT2019} employ singular value decomposition (SVD) to transform the MIMO channel into a set of subchannels. The achievable region is then characterized in terms of single-letter mutual informations optimized over all possible ADC and power allocations among the subchannels. 

In this paper, recognizing the high complexity of the optimal ADC allocation scheme for the adaptive threshold receiver \cite{abbasMtISIT2019}, we compare various low-complexity algorithms for transmit power and ADC allocation among subchannels taking into account practical constraints such as limited modulation levels and realistic mmWave channel models. We show through simulations that simple power and ADC allocation strategies are able to achieve near optimal rates for PtP communication in practical mmWave cellular networks. Additionally, we demonstrate that with the adaptive threshold receiver, using few one-bit ADCs is enough to achieve near optimal performance in terms of throughput.
In addition, we relax the idealistic assumption made in \cite{abbasMtISIT2019} that both transmitter and receiver have access to full channel state information (CSI), and consider practical channel estimation using low resolution ADCs configured with the adaptive threshold receiver. We note that prior works have considered channel estimation
when analog-one-shot receivers are used \cite{mezghani2010multiple,zeitler2012bayesian,dab2010ches,mo2018channel,shlezinger2018asymptotic}.
We show that in practical DL mmWave communication scenarios with imperfect CSI and limited number of one-bit ADCs, the achievable rate distribution is close to the one with perfect CSI and fully digital receiver with high resolution ADCs employing time division multiple access (TDMA) of equal time-shares.  
We demonstrate that the performance of the proposed adaptive threshold based TDMA in \cite{abbasMtISIT2019} outperforms that of the conventional TDMA in terms of the system throughput significantly.

\begin{figure*}[t]
\centering
\includegraphics[width =0.55\textwidth ,draft=false]{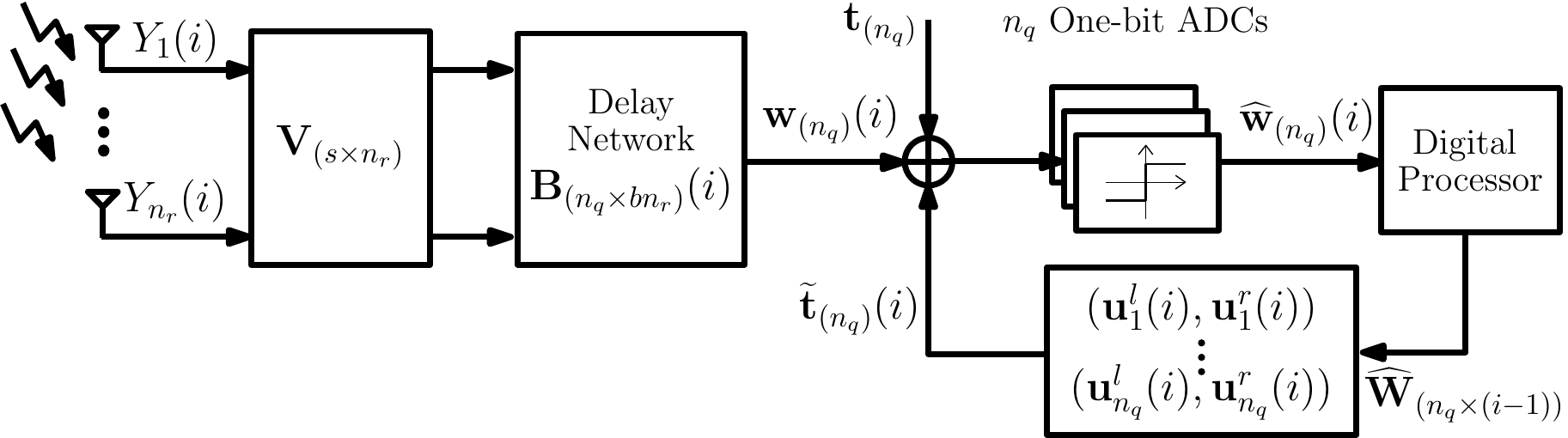}
\caption{An adaptive threshold receiver with $n_q$ one-bit ADCs is shown where the analog linear combiner, the delay network operation, and adaptive threshold coefficient vector set at channel-use $i$ are characterized by the matrix $\textbf{V}$, binary matrix $\textbf{B}(i)$, and set $\{(\textbf{u}^l_{1}(i),\textbf{u}^r_{1}(i)), \cdots, (\textbf{u}^l_{n_q}(i),\textbf{u}^r_{n_q}(i))\}$, respectively.}
\label{fig:PtP}
\vspace*{-0.4cm}
\end{figure*}

\vspace*{-0.2cm}
\section{System Model and Preliminaries}
\vspace*{-0.2cm}
\label{sec:System Model}
\subsection{Channel Model}
We consider the DL communication in a mmWave single-cell system consisting of one base station (BS) and $n_u$ users, where the $i$th user is equipped with $n_{q,i}$ one-bit ADCs. The received signal at the $i$th user is represented by
\begin{align}
\label{eq:channel}
    \textbf{y}_{i} = \textbf{H}_{i}\textbf{x} + \textbf{z}_{i},
\end{align}
where $\textbf{x}\in \mathbb{C}^{n_b}$ is the vector of transmit signal from the BS with $E[||\textbf{x}||^2] \leq P$, where $P$ is the average transmit power of the BS, $\textbf{y}_i$ is the vector of the received signal at the $i$th user, $\textbf{z}_i$ is a vector of independent, zero-mean and unit-variance complex Gaussian noise, and $\textbf{H}_i\in \mathbb{C}^{ n_{i}\times n_b}$ is the complex channel gain matrix between the BS and $i$th user, where $n_i$ and $n_b$ are the number of antennas at the $i$th user and BS, respectively. 

To model the channel in mmWave bands, we adopt a standard multipath clustered channel model described in \cite{akdeniz2014millimeter}. To elaborate, the channel between the BS (with $n_t$ antennas) and a user (with $n_r$ antennas) consisting of $n_c$ clusters (also called paths), where the $j$th cluster includes $n_{\textrm{p},j}$ rays (or sub-paths) is defined as follows
\begin{align}
\begin{aligned} 
\textbf{H} = \sum_{j=1}^{n_{c}} \sum_{k=1}^{n_{\textrm{p,j}}} \beta \times g_{j, k} \times \textbf{a$^{\textrm{r}}$}(&\varphi^{\textrm{r}}_{j, k}, \theta^{\textrm{r}}_{j, k}) 
 \textbf{a$^{\textrm{t}}$}(\varphi^{\textrm{t}}_{j, k}, \theta^{\textrm{t}}_{j, k})^*,
\end{aligned}
\end{align}
where $\beta$ denotes the large-scale fading coefficient modeling distance-dependent path loss and shadowing and $g_{j,k}$ is the complex small-scale fading gain. Also, $\textbf{a$^{\textrm{r}}$} \in \mathbb{C}^{n_r}$ and $\textbf{a$^{\textrm{t}}$}\in \mathbb{C}^{n_t}$ denote the array response vectors of the user and BS, respectively.
Furthermore, $\varphi^{\textrm{r}}_{j,k}$, $ \theta^{\textrm{r}}_{j,k}$, $\varphi^{\textrm{t}}_{j,k}$, and $\theta^{\textrm{t}}_{j,k}$ are the azimuth angle of arrival (AoA), elevation AoA, azimuth angle of departure (AoD), and elevation AoD associated with the $k$th ray of the $j$th cluster, respectively.

\vspace*{-0.2cm}
\subsection{Receiver Architecture}
\vspace*{-0.1cm}
For each user equipped with $n_r$ antennas and $n_q$ one-bit ADCs, we use the adaptive threshold receiver proposed in \cite{abbasMtISIT2019}. Here, we provide a brief description of this receiver and refer the reader to \cite{abbasMtISIT2019} for more details. The block diagram is shown in Fig. \ref{fig:PtP}. In this receiver, the output of the ADCs at the $i$th channel-use is
\begin{align}
\label{eq:ADC_out}
\widehat{\textbf{w}}(i) = Q(\textbf{w}(i)+\widetilde{\textbf{t}}(i)+\textbf{t}),
\end{align}
where $\textbf{w}(i) = \textbf{B}_{(n_q \times b n_r)}(i) \widetilde{\textbf{y}}(i)$ 
holds $n_q$ elements of the analog signal vector $\widetilde{\textbf{y}}(i)$ that are selected using the binary matrix $\textbf{B}(i)$ and are fed to the ADCs in the $i$th channel-use.
The vector $\widetilde{\textbf{y}}(i) = (\widehat{\textbf{y}}^T(b k+1), \widehat{\textbf{y}}^T(b k+2),\cdots,\widehat{\textbf{y}}^T(b (k+1))^T$ where $\widehat{\textbf{y}}(i)=\textbf{V}\textbf{y}(i), k =  \textrm{mod}_b(i)-1$ represents the concatenation of $b$ consecutive channel outputs which are processed through the linear analog combiner matrix $\textbf{V}$. These channel outputs are buffered in the delay network, and are jointly processed in analog domain at the receiver.
The vector $\tilde{\textbf{t}}$ represents the adaptive part of the ADC thresholds whose elements follow the equation
\begin{align}
\tilde{t}_k(i) = \textbf{u}^l_{k}(i)\widehat{\textbf{W}}\textbf{u}^r_{k}(i), \text{ for } k \in[n_q].
\end{align}
The matrix $\widehat{\textbf{W}}(i) = [\widehat{\textbf{w}}(kb+1), \widehat{\textbf{w}}(kb+2)\cdots, \widehat{\textbf{w}}(i-1)]$ represents the ADC outputs from the channel-uses $kb + 1$ to $i-1$, where $k = \textrm{mod}_{b}(i)$. 
The vector set $\{ (\textbf{u}^l_{1}(i),\textbf{u}^r_{1}(i)),(\textbf{u}^l_{2}(i),\textbf{u}^r_{2}(i)), \cdots, (\textbf{u}^l_{n_q}(i)$ $,\textbf{u}^r_{n_q}(i))\}$ is called the \textit{adaptive threshold coefficient vector set} and denotes the linear rule which determines the threshold of the ADCs at the $i$th channel-use with respect to the ADC outputs in the previous channel uses. The vector $\textbf{t}$ in Equation \eqref{eq:ADC_out} represents the fixed part of the ADC thresholds.
In \cite{abbasMtISIT2019}, it is shown that the adaptive threshold receiver allows for $n_q$-bit quantization using $n_q$ one-bit ADCs where the $n_q$ bits can be allocated to the antennas.
Furthermore, it is proved that in high SNR regime, this receiver achieves the transmission rate of $n_q$ bits per channel-use which is optimal among all the receivers with the same number of one-bit ADCs. Moreover, this optimal rate is achieved using practical modulation schemes such as pulse amplitude modulation (PAM) and quadratic amplitude modulation (QAM). This is explained in more detail in Section \ref{subsec:PtP}.

\vspace*{-0.2cm}
\section{Communication Schemes and Channel Estimation}
\label{sec:PtP}
In this section, we first consider communication over a PtP MIMO system with adaptive threshold receiver and perfect CSI available at both transmitter and receiver terminals. Then, we investigate a DL scenario. We further investigate channel estimation using one-bit ADCs and the impact of imperfect CSI on the proposed schemes. 

\vspace*{-0.2cm}
\subsection{PtP and DL Communication with Perfect CSI}
\label{subsec:PtP}
\textbf{PtP communication:} Consider a PtP MIMO system with the adaptive threshold receiver, where the transmitter is equipped with $n_t$ antennas, the receiver is equipped with $n_r$ antennas and $n_q$ one-bit ADCs, and the channel is represented as in Equation \eqref{eq:channel} with channel matrix \textbf{H}. It is assumed that both transmitter and receiver terminals have perfect CSI. We consider the communication scheme described in {\cite[Theorem 1]{abbasMtISIT2019}} which is summarized in the following. 
In the first step, singular value decomposition (SVD) is performed in the analog domain to transform the complex channel \textbf{H} into $s$ parallel real subchannels. Let $\sigma_{k},~k \in [s]$  represent the singular values associated with the per dimension of the channel gain matrix  ${\textbf{H}}$ (i.e., $\textrm{Re} \left (\textbf{H}\right)$ and $\textrm{Im} \left (\textbf{H}\right) $). Fix $n_{q,1}, n_{q,2}, \cdots, n_{q,s}\in \mathbb{N}\cup \{0\}$ and $P_1,P_2,\cdots,P_s\in \{\mathbb{R}^{+},0\}$ such that $\sum_{k\in [s]} n_{q,k}=n_q$ and $\sum_{k\in [s]}P_k=P$, where $n_{q,k}$ and $P_k$ are the number of one-bit ADCs and transmit power allocated to the $k$th subchannel, respectively. In Section \ref{sec:numerical_results}, we provide several low-complexity algorithms for power and ADC allocation and compare their performances in terms of achievable rates under realistic channel models. The matrices \textbf{V} and \textbf{B}, and the adaptive threshold coefficient vector set are taken so as to ensure that the $n_{q,k}$ one-bit ADCs allocated to the $k$th subchannel perform as $n_{q,k}$-bit quantization following \cite{abbasMtISIT2019}. The transmitter uses $2^{n_{q,k}}$-PAM signaling for each real subchannel. An example of this receiver is provided below.
\vspace*{-0.1cm}
\begin{Example}
For a PtP single-input single-output (SISO) system with channel gain $\textbf{H} = 1$, and $n_q= 2$, a possible set of values of the receiver parameters are as follows: $\textbf{V} = 1$, $\textbf{u}^l_j = \textbf{1}_j, j\in\{1,2\}$
\begin{gather} 
\textbf{t} = \begin{bmatrix}0 \\ 0 \end{bmatrix},
 \textbf{B} = \begin{bmatrix}1 & 0 \\ 0 & 1 \end{bmatrix},\textbf{u}^r_1(i) = \textbf{u}^r_2(i) = \begin{cases} 0 & \text{$i$ is even}\\
 -\frac{1}{2} & \text{$i$ is odd}
  \end{cases},
\end{gather}
where $\textbf{1}_j$ is the indicator vector whose $j$th element is one and the rest are zero. To elaborate on how this receiver operates, let us consider the first three channel-uses. After two channel-uses, the thresholds of ADCs are set to zero and the channel outputs in the first and second channel-uses are fed to the first and second ADC, respectively. In the third channel-use, the thresholds of ADCs are set to half of their outputs in the previous channel-use and a delayed version of the first two channel outputs are fed to the ADCs. The receiver operates in a similar manner to the second and third channel-uses for the rest of the communication. 
\end{Example}

\noindent\textbf{DL communication: }
In DL communication, a TDMA protocol is used where users are scheduled in a round robin fashion so that the BS transmits to one user at each channel-use. We assume the BS transmits to each user once every $n_u$ channel-uses. Each user activates its adaptive threshold receiver for all of the $n_u$ channel-uses, i.e the user's ADCs are active even when the BS does not transmit to it \cite{abbasMtISIT2019}. As in the PtP scenario, SVD, power and ADC allocation across subchannels are performed for optimizing communication rates. Let $n_{q,i,k}$ denote the number of one-bit ADCs allocated to the $k$th subchannel of the $i$th user. The BS uses $2^{n_u n_{q,i,k}}$-PAM signaling over that subchannel to compensate for the $n_u-1$ channel-uses it does not send information over that subchannel.
Compared to a naive time-sharing strategy where a user's receiver is only active when the BS transmits to that user, the proposed TDMA strategy increases the communication rate. 
Under the proposed scheme, at high SNR, each user achieves the optimal transmission rate which is $n_{q,i}$ bit per channel-use for the $i$th user. 
In contrast, the naive TDMA strategy leads to the high SNR achievable rate of $n_{q,i}/{n_u}$ bit per channel-use.
We elaborate more on this in Section \ref{sec:numerical_results}. An example of the receiver parameters is provided below.
\begin{table}[t]
\centering
\label{tab:sim-param}
\caption{Simulation Parameters}
\begin{footnotesize}
\begin{tabular}{ll} \toprule
$\bf Parameter$& $\bf Value$\\ \midrule
Cell radius& $10$ to $50$ m\\
Carrier frequency& $28$ GHz\\
Bandwidth& $1$ GHz\\
Noise spectral density& $-174$ dBm/Hz\\ 
Noise figure& $6$ dB\\ 
BS antenna & $8$x$8$ uniform planar array\\
User antenna & $4$x$4$ uniform planar array\\
BS transmit power & $30$ dBm\\
Path loss (LOS) in dB & $61.4+20\log_{10}(d~\text{in m}) + {\cal{N}}(0, 5.8^2)$\\
Path loss (NLOS) in dB & $72+29.2\log_{10}(d~\text{in m}) + {\cal{N}}(0, 8.7^2)$\\
Probability of  LOS & $\textrm{exp}(-0.0149d~\text{in m})$\\
\bottomrule\\
\end{tabular}
\end{footnotesize}
\vspace*{-0.4cm}
\end{table}

\vspace*{-0.2cm}
\subsection{Channel Estimation}
\label{subsec:ch}
While Section \ref{subsec:PtP} considers perfect CSI, the effects of one-bit ADCs on accuracy of channel estimation must also be taken into account. We assume that at the beginning of each coherence interval, the users perform channel estimation using expectation maximization generalized approximate massage passing (EM-GAMP) algorithm investigated in \cite{mo2018channel} which exploits the sparsity of the angular representation of the mmWave channel to reduce the estimation overhead and enhance the quality of the channel estimate. In \cite{mo2018channel}, the EM-GAMP algorithm for mmWave channel estimation with fully digital receivers equipped with a $m$-bit ADC per each dimension (real and imaginary) of the receiver antennas is studied. \textcolor{black}{Furthermore, the computational complexity of the algorithm is analyzed and methods for complexity reduction are provided.}  
Since we assume no prior knowledge about the channel such as long-term statistics, the best worst case allocation of the one-bit ADCs among antennas is the uniform one. 
Therefore, to perform channel estimation, we configure the adaptive threshold receiver for the $i$th user to form a fully digital receiver with a $m_i$-bit ADC per each dimension of each antenna, where $m_i = n_{q,i}/{2n_i}$. 
After channel estimation, the users send their CSI to the BS. Once the channel estimation step is completed, each user is configured for DL transmission as described in Section \ref{subsec:PtP} using the estimated CSI. In Section \ref{sec:numerical_results}, we demonstrate through several simulations of practical scenarios that using estimated channel through this method does not lead to significant rate-loss when the adaptive threshold receiver is used. 

\vspace*{-0.2cm} 
\section{Simulation Results}
\vspace*{-0.2cm} 
\label{sec:numerical_results}
In this section, we provide various simulations to establish the performance of the proposed architectures in Section \ref{sec:PtP}
in practical scenarios. We consider a small-cell scenario with three dimensional network model consisting of a BS and ten users operating at 28 GHz. Users are distributed uniformly in a ring around the BS with inner and outer radii of $10$ to $50$ meters. The maximum transmit power of the users and the BS are set to $23$ dBm and $30$ dBm, respectively. We consider the clustered channel model described in Section \ref{sec:System Model}. The value of the parameters in this model is adopted from \cite{akdeniz2014millimeter} where an empirical approach is taken to estimate these parameters. To elaborate, we assume that the number of clusters in the channel of each user follows a Poisson-max distribution (i.e. $n_{c,j}=\max\{1,Poisson(\lambda)\}$) with mean $\lambda=1.8$ and $20$ rays per cluster. Furthermore, we assume that the BS and users are equipped with $8\times 8$ and $4\times 4$ uniform rectangular antenna arrays, respectively. We consider a maximum spectral efficiency of 8 bps/Hz which is equivalent to use of 256 QAM modulation (i.e. 16 PAM per real dimension) as envisioned for 5G NR standard \cite{5G-NR}. We assume that the coherence time of the channel is $n_c = 10240$ as in \cite{mo2018channel}.
Table I 
lists the details of  the simulation parameters chosen.

\begin{figure}[t]
    \centering
    \includegraphics[width  = 0.45\textwidth]{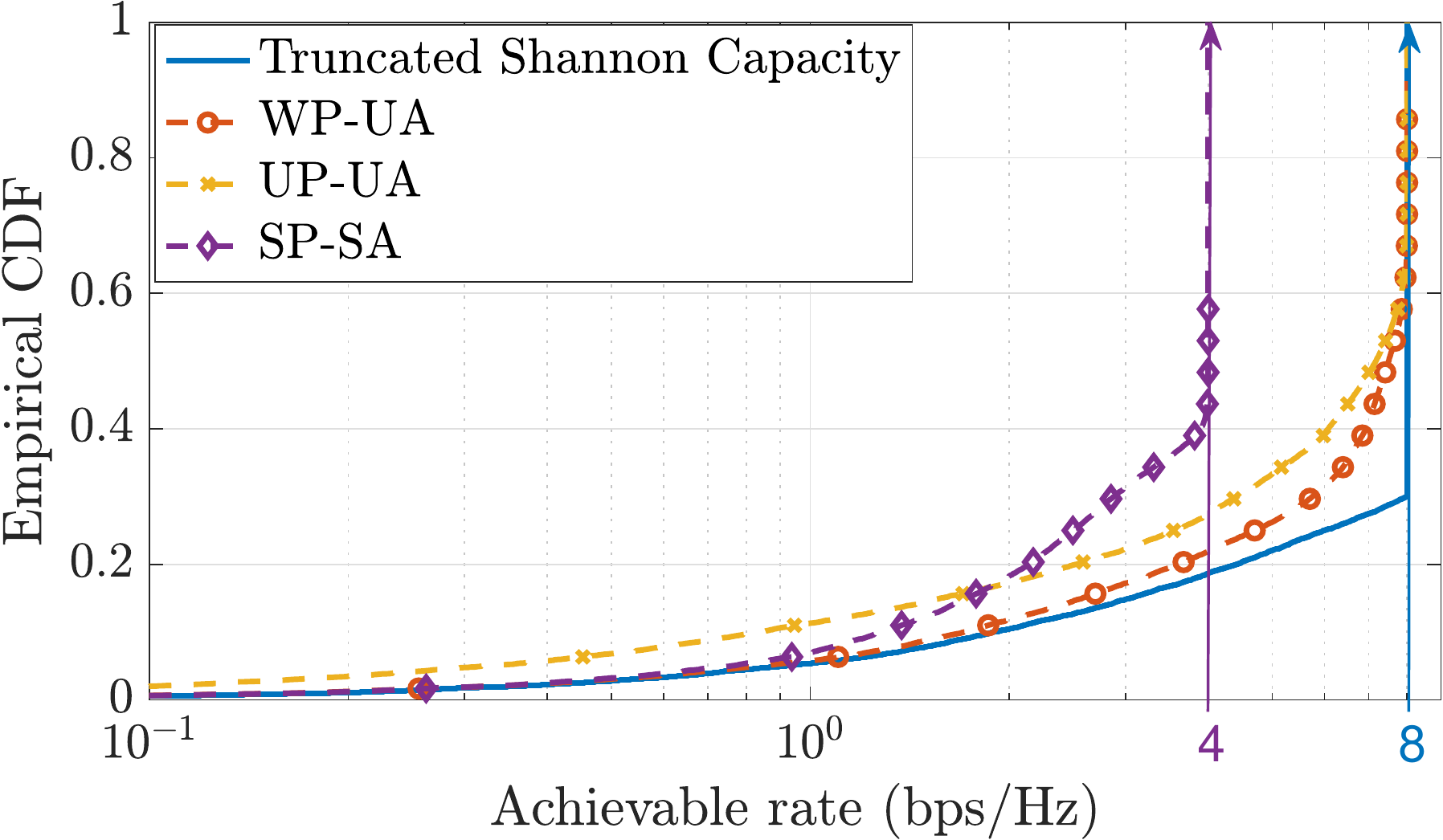}
    \caption{Empirical CDF of the achievable rate of the PtP system with perfect CSI when the receiver is equipped with $n_q = 8$ one-bit ADCs.}
    \label{fig:PtP_1}
    \vspace*{-0.4cm}
\end{figure}

\vspace*{-0.2cm}
\subsection{Power and ADC Allocation with Perfect CSI}
The transmission schemes described in Section \ref{subsec:PtP} use SVD to transform the MIMO channel into a set of parallel subchannels and then distribute the transmit power and ADCs among them. Finding the optimal distribution of the transmit power and ADCs is equivalent to solving a mixed integer programming problem which is known to be NP-hard \cite{bixby2004mixed}. Here, we investigate the performance of several practical heuristic power and ADC allocation approaches. To this end, in this section we consider a PtP scenario with perfect CSI and
$n_q = 8$. We compare the achievable rate for the following power and ADC allocation strategies: \\
    $\bullet$ \textbf{WP-UA (Waterfilling Power/Uniform ADCs):} This heuristic employs waterfilling \cite{cover2012elements} for power allocation among subchannels and assigns the ADCs to each subchannel uniformly. Note that this may result in a non-uniform ADC assignment to receive antennas.\\
    $\bullet$ \textbf{UP-UA (Uniform Power/Uniform ADCs):} In this approach, both the transmit power at the transmitter and ADCs at the receiver are distributed uniformly among the subchannels.\\
    $\bullet$ \textbf{SP-SA (Selection Diversity):} This approach allocates all the power and ADCs to the strongest subchannel.

Fig. \ref{fig:PtP_1} illustrates the empirical cumulative distribution function (CDF) of the achievable rates for the described power and ADC allocation methods for $n_q = 8$. As a performance benchmark and an upperbound for the achievable rates, we consider the truncated Shannon capacity of the MIMO channel which is $\min\{ C,n_q\}$, where $C$ is the Shannon capacity. Fig. \ref{fig:PtP_1} suggests that WP-UA has better performance compared to the other approaches while in the power-limited regime (low rates) SP-SA approach achieves comparable performance.
We also note that the modulation cap restricts the performance of the SP-SA approach. Furthermore, we can see that the WP-UA heuristic performs close to the truncated Shannon upperbound implying that using complex optimization for joint power and ADC allocation would only lead to incremental improvements in the achievable rates. We note that for users with low and intermediate SNRs (for which the achieved data rates are up to $4$ bps/Hz), as well as for users with high SNRs using only a few one-bit ADCs (e.g. one or two one-bit ADCs per real subchannel as in Fig. \ref{fig:PtP_1}) leads to a near-optimal data rates.
\vspace*{-0.3cm}
\subsection{Impact of Imperfect CSI}
\vspace*{-0.1cm}
In this section, we investigate the impact of imperfect channel estimation on the performance of the proposed architectures for PtP and DL scenarios described in Section \ref{subsec:PtP}.
To estimate the channel matrix of each user, we proceed as explained in Section \ref{subsec:ch}. The BS transmits a pilot sequence of length $n_p = 512$ and the users perform channel estimation using three one-bit ADCs per dimension (real and imaginary) of each antenna which are configured as a $3$-bit ADC using the adaptive threshold receiver. Note that using three one-bit ADCs instead of a $3$-bit ADC can potentially reduce the power consumption at the receiver. One possible venue for future work is to use knowledge of the long-term channel statistics to reduce the number of required ADCs while achieving similar performance in channel estimation.

\begin{figure}[t]
    \centering
    \includegraphics[width  = 0.45\textwidth]{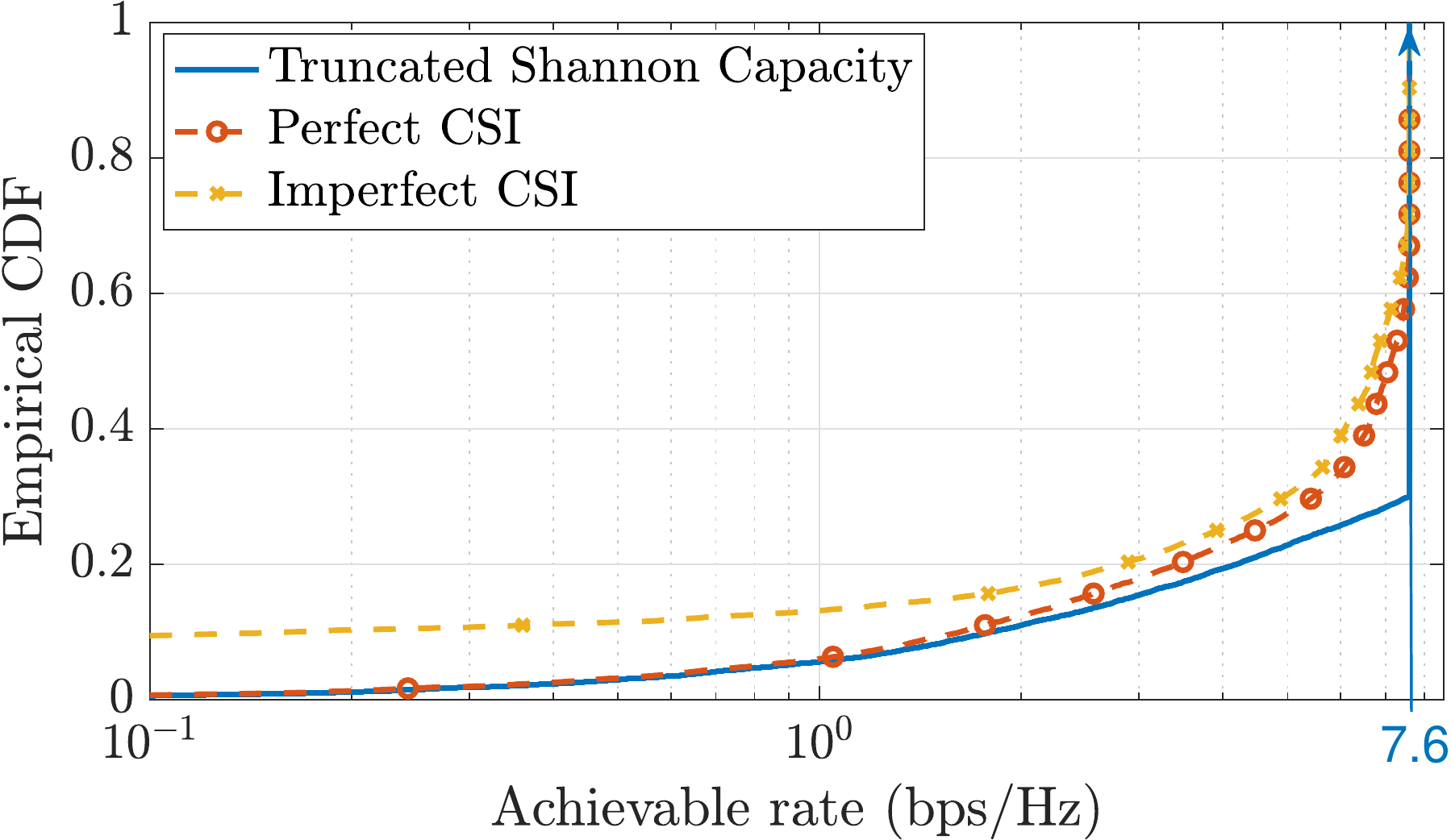}
    \caption{Empirical CDF of the achievable rate of the PtP system with $n_q=  8$ one-bit ADCs at the receiver.}
    \label{fig:MAC}
    \vspace*{-0.5cm}
\end{figure}

Although we use three one-bit ADCs per dimension ($16\times 2\times 3 =96$ in total) during channel estimation, motivated by Fig. \ref{fig:PtP_1}, we do not need that many ADCs to achieve near-optimal performance during data transmission. The reason is that we can use CSI (available after channel estimation) to exploit the sparsity of the channel. Moreover, using fewer ADCs leads to lower power consumption during the data transmission phase. Therefore, we assume that the users only use $n_{q,i} = 8$ one-bit ADCs during data transmission. Also, for ADC and power allocation we use the WP-UA heuristic. \textcolor{black}{To calculate the achievable rate of the system, we design the system parameters such as modulation points, analog linear combiner as described in Section \ref{subsec:PtP} using the estimated channel. Next, we numerically calculate the transmission probability matrix of the corresponding discrete input discrete output system. Then, to determine the achievable rate, we compute the mutual information given a uniform prior for input.}

The empirical CDF of the PtP achievable rates with perfect and imperfect CSI is illustrated in Fig. \ref{fig:MAC}. As an upper bound, we use $\frac{n_c-n_p}{n_c}\min\{ C, n_q  \}$, where $C$ is the Shannon capacity with perfect CSI and high resolution ADCs. We note that in the presence of the channel estimation error, the MIMO subchannels after performing SVD will interfere with each other which degrades the performance. Comparing the CDF of the achievable rates with perfect and imperfect CSI in Fig. \ref{fig:MAC}, we observe that while the performance loss is small for intermediate and high SNRs, it is larger in the low SNR regime. This is due to the fact that in low SNRs the channel estimation error is high. 

\begin{figure}[t]
    \centering
    \includegraphics[width  = 0.45\textwidth]{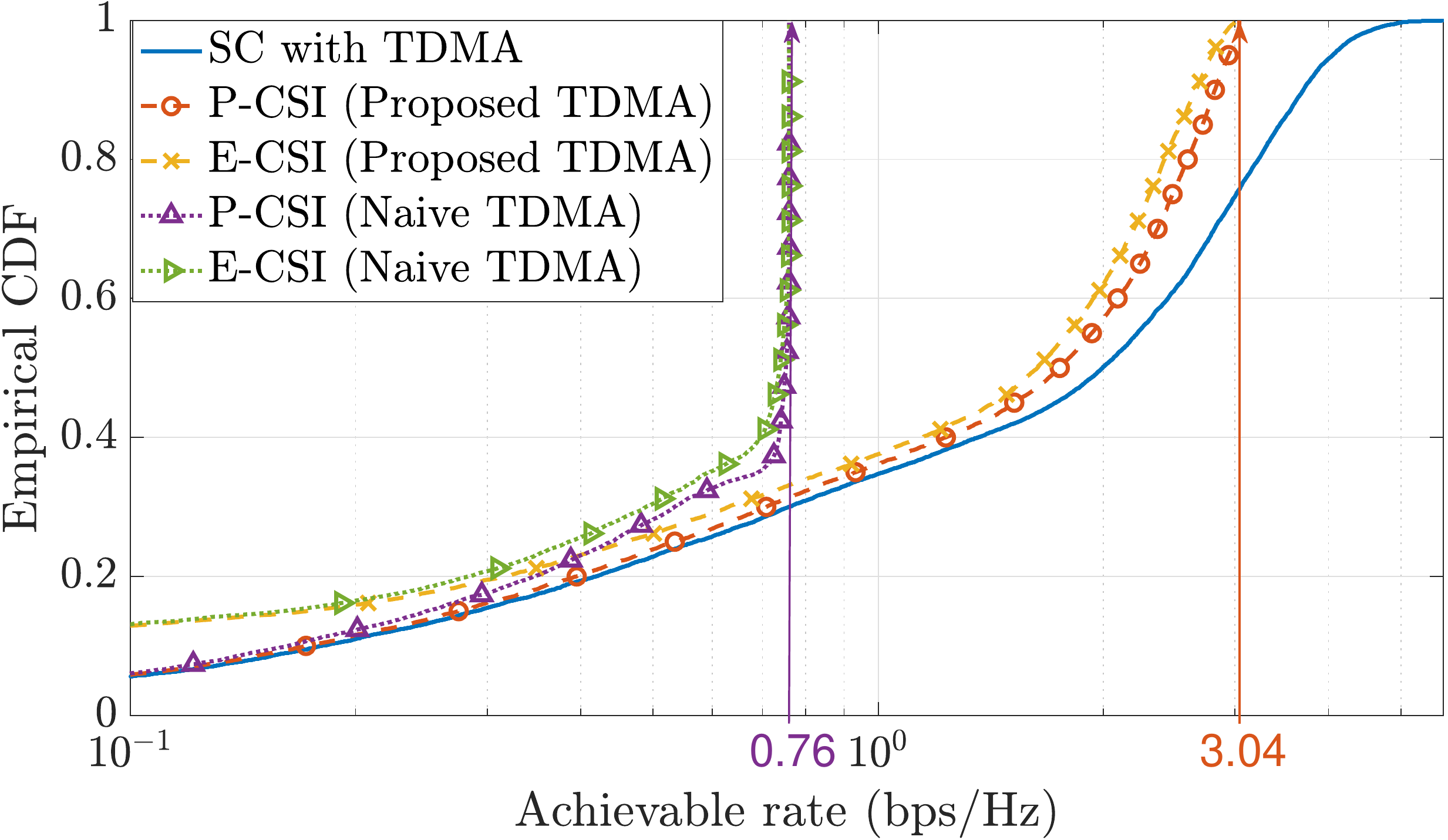}
	    \caption{Empirical CDF of the achievable rates of the users for DL transmission when $n_u = 10$ and the users are equipped with $n_{q,i}=8$ one-bit ADCs. SC, P-CSI, and E-CSI denote Shannon capacity, perfect CSI, and estimated CSI, respectively. }
    \label{fig:BC}
    \vspace*{-0.5cm}
\end{figure}

Fig. \ref{fig:BC} depicts the empirical CDF of the DL per user achievable rates with perfect and imperfect CSI. We use $ \frac{n_c-n_p}{n_c}\min\{C_{t}, n_{q,i}\}$, where $C_t$ is the Shannon capacity with TDMA of equal time-shares in the presence of perfect CSI and high resolution ADCs as an upperbound. Since $n_{q,i} = 8$, the truncation effect cannot be observed in the range of the plot. We observe that the performance loss caused by the estimation error is small in intermediate and high SNR regime. Furthermore, we see that the proposed TDMA approach for DL with adaptive threshold receiver discussed in Section \ref{subsec:PtP} provides a performance close to the upperbound. Note that, as discussed in Section \ref{subsec:PtP}, while the proposed TDMA strategy leads to higher power consumption at the users compared to naive TDMA since their ADCs are active in all the channel-uses, it significantly increases the system's achievable rate (up to $4\times$ over naive TDMA).

\vspace*{-0.3cm}
\section{Conclusion}
\vspace*{-0.2cm}
\label{sec:conclusion}
In this paper, we have considered energy efficient multiuser communication in a practical mmWave DL scenario, where the receiver is equipped with one-bit ADCs. We have compared low-complexity algorithms for the power and ADC allocation among transmitter and receiver terminals, respectively. We have shown that under practical mmWave settings with limits on the modulation levels, and imperfect channel estimation, using low resolution ADCs with the adaptive threshold receiver under simple allocation algorithms does not notably degrade the performance in terms of achievable rates. 
We have provided simulations for multiuser DL communication scenarios which show that the achievable rate of the proposed architecture is close to the optimal Shannon rate when using TDMA protocol where users have equal time-shares with high resolution ADCs.

\bibliographystyle{IEEEbib}

\bibliography{ref}

\end{document}